# Experimental Quantum Error-Free Transmission

Xian-Min Jin[1*], Zhen-Huan Yi[2], Bin Yang[1], Fei Zhou[2], Tao Yang[1] and Cheng-Zhi Peng[1,2*]

[1]Hefei National Laboratory for Physical Sciences at Microscale and Department of Modern Physics, University of Science and Technology of China, Hefei, Anhui, 230026, PR China

[2]Physics Department, Tsinghua University, Beijing 100084, PR China

*e-mail: jinxm@ustc.edu.cn; pcz@mail.tsinghua.edu.cn

**Error-free transmission (EFT) of quantum information is a crucial ingredient in quantum communication network[1]. To overcome the unavoidable decoherence in noisy channel, to date, many efforts have focused on faithfully transmitting one state by consuming large numbers of synchronized ancillary states. However, huge demands of quantum resources are hard to meet with current technology, thus restrict practical applications. Here we propose and demonstrate an economical method of reliably transmitting quantum information. An arbitrary unknown quantum state is converted into time bins deterministically in terms of its own polarization using a modified Franson interferometer[2]. Any arisen noise in channel will induce an associated error to the reference frame of the time bins, which can be utilized to reject errors and recover the initial state. By virtue of state-independent**

**feature, our method can be applied to entanglement distribution. After passing through 0.8 km randomly twisted optical fiber, the entanglement still survives and is verified. Our approach significantly simplifies the implementation of quantum EFT and enables a general quantum state even entanglement to be protected by feedback, thus can be used as a basic building block in practical long-distance quantum communication network.**

Quantum error correction was first proposed to protect a fragile quantum state by encoding the state into a multi-particle entangled state[3, 4, 5, 6]. The errors in the evolution of a quantum system can be detected and corrected. So far, significant experimental progresses have been achieved in the nuclear magnetic resonance[7, 8, 9], ion-trap[10] and linear optics[11]. Another original approach is quantum purification[12, 13, 14], which can extract a certain number of almost perfectly pure states out of a large number of less-pure states using local operation and classical communication. Besides, some efforts have been made to transmit a quantum state against a specified error involving less number of ancillary particles, such as schemes employing decoherence-free subspace for collective noise[15, 16, 17, 18, 19], quantum error rejection for bit-flip error (BFE)[20, 21] and quantum error filtration for phase-shift error (PSE)[22, 23]. However, it remains an experimental challenge to implement quantum EFT in long distance. The main practical obstacle is that required large number of single-photon or entangled sources need to be deterministic or synchronized. This is far beyond what are experimentally possible at present. In this letter, we propose and demonstrate an economical EFT scheme with the assistance of reference frame correction.

Without loss of generality, let us consider the scenario that Alice wants to send an arbitrary polarization state $|\psi\rangle = \alpha|H\rangle + \beta|V\rangle$ to Bob via a noisy quantum channel, where $\alpha$ and $\beta$ are two complex numbers satisfying $|\alpha|^2 + |\beta|^2 = 1$, $|H\rangle$ ($|V\rangle$) denotes horizontally (vertically) linear polarization. As shown in Fig.1a, instead of directly sending the initial state to Bob, Alice encodes it into two time bins with specified polarization using an unbalanced Mach-Zehnder interferometer (UMZI). The configuration of two polarization beam splitters (PBS) enables H mode to go short path and V mode to go long path in a deterministic way:

$$\hat{T}|\psi\rangle = \alpha|H\rangle + \beta|V\rangle_T \qquad (1),$$

where subscripts $T$ denotes additional time delay on $V$ mode induced by UMZI$_1$. A general error in noisy channel can be expressed as $E = e_0 I + e_1 X + e_2 Z + e_3 XZ$ ($e_0^2 + e_1^2 + e_2^2 + e_3^2 = 1$), where three Pauli matrices $I = \begin{pmatrix} 1 & 0 \\ 0 & 1 \end{pmatrix}$, $X = \begin{pmatrix} 0 & 1 \\ 1 & 0 \end{pmatrix}$ and $Z = \begin{pmatrix} 1 & 0 \\ 0 & -1 \end{pmatrix}$ denote identity, BFE and PSE respectively[24]. At Bob's station, he applies a time delay $T'$ on $H$ mode with UMZI$_2$. The output state with specified error can be written as:

$$\hat{T}'I\hat{T}|\psi\rangle = \alpha|H\rangle_{T'} + e^{i\phi_1}\beta|V\rangle_T \qquad (2a),$$

$$\hat{T}'X\hat{T}|\psi\rangle = \alpha|V\rangle + e^{i\phi_2}\beta|H\rangle_{TT'} \qquad (2b),$$

$$\hat{T}'Z\hat{T}|\psi\rangle = \alpha|H\rangle_{T'} - e^{i\phi_1}\beta|V\rangle_T \qquad (2c),$$

$$\hat{T}'XZ\hat{T}|\psi\rangle = \alpha|V\rangle - e^{i\phi_2}\beta|H\rangle_{TT'} \qquad (2d),$$

where $\phi_1 = kc(T - T')$, $\phi_2 = kc(T + T')$, $k$ is the wave vector and $c$ is the velocity of photon in two UMZIs.

Obviously, the two time bins in equation (2b) or (2d) are so separate sufficiently (longer than coherence time of state-carrier photon (SP)) that they decoherence to be $\alpha^2 |V\rangle\langle V| + \beta^2 |H\rangle_{TT'}\langle H|$, thus become two side peaks in Fig1.b. The two time bins in equation (2a) or (2c) overlap in UMZI$_2$ again and settle in the central peak. The BFE and bit-phase-flip errors can be readily rejected by post-selecting the central peak and discarding the events emerged in the two side peaks. This step can be realized by comparing time interval of sending and receiving the state with the assistance of synchronization clocks. If the interval equals to $\Delta t + T$ (central peak), where $\Delta t$ is the required time that SP takes flying from Alice to Bob, Bob can ensure the received state has no BFE. If the interval is $\Delta t$ or $\Delta t + 2T$ (side peaks), it is obvious that BFE has occurred during the transmission. Bob just needs to discards the state. This process can completely reject BFE meanwhile reduce transmission efficiency to $e_0^2 + e_2^2$.

In our special EFT system, PSE behaves as the time drift between two bins. A weak coherent laser with well-defined polarization $|+\rangle = \frac{1}{\sqrt{2}}(|H\rangle + |V\rangle)$ is employed to probe the time drift by coupling in and out of EFT system. The probe process is independent of transmitting states. Therefore, it can run continuously to give an offset to a feed-back controller, which is used to lock the time drift at 0. If the PSE happens as equation (2c), our reference frame correction can be utilized to correct it by adjusting the locked phase to be $-e^{i\phi_1}=1$. Finally, Bob can ensure that every retrieved quantum state is exactly same with the state originally sent by Alice.

In the experiment, type-II spontaneous parametric down conversion (SPDC)[25] is employed to generate a heralded single photon as SP [See Method]. As is shown in Fig.1a, an arbitrary polarization state $|\psi\rangle = \alpha|H\rangle + \beta|V\rangle$ is prepared by a set of optical components consisting of a polarizer, a half wave plate (HWP) and a quarter wave plate (QWP). The time delay $T$ introduced by UMZI$_1$ is about 2.5ns. The photon subsequently passes a polarization controller (PC), two pieces of 5m single mode fiber (SMF), and a BFE simulator. The special configuration of PC, a HWP sandwiched by two QWPs, has been proved capable of unitarily realizing arbitrary polarization manipulation. Hence it can be utilized to initialize the error of whole channel. A further testing for longer SMF can be realized by simply inserting 0.8km fiber. In order to see the performance of EFT system, a noise simulator (NS) is necessary to give the quantified BFE in the noisy channel. Its structure is identical to PC but with fixed setting for each QWP at $90°$. By randomly setting the HWP axis to be oriented at $\pm\theta$, the noisy quantum channel[21] can be engineered with a BFE probability of $Sin^2(2\theta)$. UMZI$_2$ subsequently applies 2.5 ns time delay $T'$ on the $H$ mode of the state by guiding photon into the interferometer from the other input port, see Fig.1a. To see time profile of output states, we scan the detected signal of trigger photon (TP) with a range of 16 ns to see 2-fold coincidence rates. We choose to observe two representative cases of 0% and 100% BFE rates by setting HWP in NS at $0°$ and $45°$ respectively. We obtain three expected peaks respectively corresponding to the interval of $\Delta t$, $\Delta t + T$ and $\Delta t + 2T$. As shown in Fig.1c, experimental data illustrates that BFE can be effectively rejected by comparing the interval of TP and SP's registration on silicon avalanche photodiodes. The window of coincidence logic is set at 2ns, which is sufficient to filter the two side peaks, only the events without BFE are preserved.

To protect the coherence between two terms in equation (2a) and (2c), it is necessary to fix $\Delta T = T - T'$ to be 0, at least within the coherent length of SP ($l = \frac{\lambda^2}{\Delta \lambda}$). We firstly need to find the exact overlap position of two time bins by scanning UMZI$_2$ and measuring the output probe beam at the bases of $|+\rangle$ or $|-\rangle$. An interference envelope can be observed around the zero difference between two UMZIs, see Fig.1d. Experimentally, all the photons are collected with SMFs to define the exact spatial mode. Narrow bandwidth filter with FWHM 4 nm is set in front of each detector to define the exact spectral mode ($l = 164 \mu m$). An additional prism built in a precise linear stage is employed to achieve accurate temporal overlap for two time bins[26]. At the middle of the envelope, by scanning with a step size of 30nm, we observed phase oscillation with high visibility, $V = (C_{peak} - C_{valley})/(C_{peak} + C_{valley}) = 99.2\%$, where $C_{peak}$ ($C_{valley}$) is the correlated coincidence rate at the peak (valley) of the two-fold coincidence curve. The high visibility implies that the spatial and temporal mode of two time bins overlap perfectly on the second PBS of UMZI$_2$.

To verify the success of EFT, without loss of generality, we select linear polarization state $|H\rangle$, $|+\rangle = \frac{1}{\sqrt{2}}(|H\rangle+|V\rangle)$ and circular polarization state $|R\rangle = \frac{1}{\sqrt{2}}(|H\rangle+i|V\rangle)$ as initial states to be transmitted. BFE rates are quantitatively introduced by changing the axis of HWP in BFE simulator. Specifically, we vary the angle $\theta$ to achieve various error rates from 0 to 1 with an increment of 0.1 in the quantum channel. We investigate the change of error rates and coincidence rates under these conditions with 10m and 0.8km fiber. As shown in Fig. 2, the errors in noisy channel increase with the angle of $\theta$ but the final errors of output states after EFT are all confined within 10%, except the case of 100% BFE. In fact, the error rates mainly come from the imperfection of state preparation and signal-to-noise ratio. In the case of inserting 0.8km fiber, the performance of EFT is very similar to the case of short fiber, representing significant robustness of our protocol. The coincidence highly depends on the channel noise. It decreases to a minimum while the error rate increases closed to 100%. By applying random BFE on the channel, the coincidences and the error rates can much less depends on the channel errors[16, 18].

The performance of the system against PSE is further tested. For the case of transmitting $|+\rangle$, we apply the feed-back control to one arm of UMZI$_2$ to correct the reference frame conditioning on the result of the time drift of two effective time bins. To achieve this, we feed a probe laser (633nm wavelength) with a well defined polarization into and out of EFT system by two dichromatic mirrors (DM, T808R633). Fig.3a shows that typical errors occurred in a twisted SMF without EFT and the errors only without active feed-back control are unacceptable for general quantum information processing. But with our EFT system, the errors are suppressed effectively to very low level. The stability and precision highly relate to the speed of feedback control.

To verify further the state-independent feature of our EFT system, and thus show the capability of distributing entanglement, we feed one of entangled photons into EFT system and a polarization analyzer in Bob side, the other one of entangled photons into a polarization analyzer in Alice side directly. The entanglement is verified by the way of measuring the violation of Clauser-Horne-Shimony-Holt (CHSH) inequality[27]. The polarization correlation function is defined as follows,

$$E(\theta_A, \theta_B) = \frac{N_{++} + N_{--} - N_{+-} - N_{-+}}{N_{++} + N_{--} - N_{+-} - N_{-+}} \qquad (3),$$

Where $N_{++}$, $N_{--}$, $N_{+-}$ and $N_{-+}$ are the coincident counts between Alice and Bob with the actual settings of $(\theta_A, \theta_B)$, $(\theta_A + \frac{\pi}{2}, \theta_B + \frac{\pi}{2})$, $(\theta_A, \theta_B + \frac{\pi}{2})$ and $(\theta_A + \frac{\pi}{2}, \theta_B)$ respectively. In the CHSH inequality, parameter S is defined as,

$$S = |E(\theta_A, \theta_B) - E(\theta_A, \theta'_B) + E(\theta'_A, \theta_B) + E(\theta'_A, \theta'_B)| \qquad (4),$$

In the local realistic view, no matter what angles $\theta_A$ and $\theta_B$ are set to, parameter S should be $S \leq 2$. But in the view of quantum mechanics, S will get to the maximal value $2\sqrt{2} \approx 2.828$ when the polarization angles are set to $(\theta_A, \theta'_A, \theta_B, \theta'_B) = (0°, 45°, 22.5°, 67.5°)$. The measurement result of S is $S = 2.70 \pm 0.05$, which violates CHSH Inequality by 14 standard deviations. This clearly confirms the quantum nature of the recovered entanglement. At 0.8km twisted SMF, we do the CHSH inequality measurement repeatedly with and without EFT system. The Fig. 3b shows that the Bell's inequality can be violated with the EFT but can not be without EFT system.

It should be noticed that the presented approach deserves further discussion. There are some adoptable improvements. For example, firstly, we can substitute polarization-maintaining fiber for normal SMF as transmitting channel. Time bins with well-defined polarization will be free of BFE completely, thus the transmission efficiency will increase substantially. One modification to EFT system is setting the value $\phi_1 = kc(T-T') + k(n_H - n_V)l$ to 0, where $l$ is the length of polarization-maintaining fiber, $n_H(n_V)$ is the refraction coefficient for $H$ ($V$) polarization. Secondly, the technology of WDM (wavelength division multiplexing) can be exploited to coupling reference frame probing laser into and out of EFT system with small loss, but the precondition is that the wavelength of probe laser should be very close to signal. Thirdly, the probe laser prepared at exactly uniform wavelength with SP can be encoded with signal in tandem, thus they can be divided by time window. The precision of probing is high related to the interval between the time bins of probe laser and followed signal, the less the interval we use, the higher precision we can get.

In summary, our EFT method with reference frame correction deserves some further comments. First, we purify arbitrary state by itself and corrected reference frame, so we don't need to provide any ancillary particle simultaneously. we can overcome BFE over the noisy channel by rejecting error directly, but meanwhile PSE can be probed in real time and correct actively. Thus efficiency of communication will be enhanced several orders even compared with the scheme involving only one ancillary particle. Second, thanks to the linear feature of quantum mechanics, our method can be generalized to transmit arbitrary multi-qubit states, e.g. GHZ states[28], Cluster states[29], W states[30] and so forth. Third, the threshold of tolerable error rate over the quantum noisy channel can be greatly improved because of rejecting BFE deterministically. The threshold is determined by signal-to-noise ratio, our scheme is always effective until dark counts are comparable to the signal counts. Finally, our EFT process is scalable with multistage in one channel and can combine with entanglement swapping[31,32] for quantum repeater[33]. These virtues make our EFT system promising for the practical task in implementation of long-distance quantum communication network.

## Methods

**Single heralded photon and entanglement preparation.** To create single photon as SP and entanglement, a semiconductor blue Laser beam (with a power of 34.5 mw, a waist of $100\,\mu m$ and a central wavelength of 405 nm) incident on a 2 mm beta-barium-borate (BBO) crystal to generate entangled photonic pairs at 810nm with type-II SPDC. The down-converted extraordinary and ordinary photons have different velocities and travel along different paths inside the crystal due to birefringent effect of the BBO crystal. The resulting walk-off effects are compensated by a combination of a HWP and an additional 1 mm BBO crystal in each arm. With single mode fibers and with 4-nm bandwidth interference filters in front of single photon detectors, we collect about 80,000 $s^{-1}$ single photons in each arm of the source and 22,000 pairs of polarization entangled photons per second. The observed visibilities of the polarization correlations, which can reveal the quality of our prepared entanglement, are about 98.1% for $|H\rangle/|V\rangle$ basis, and 92.6% for $|+\rangle/|-\rangle$ basis. In this step, by checking the entanglement visibility between the two created photons, we can get exactly identical photons in frequency with the assistance of type-II phase matching. Because SPDC creates photons in pairs, the detection of one photon indicates, or "heralds", the existence of its twin. Thus the SP can be prepared conditioning on the detection of TP

**Acknowledgements**

X.-M.J is grateful for insightful discussions with Y.-A. Chen and J.-C. Boileau. This research was supported by the Chinese Academy of Sciences, the National Natural Science Foundation of China and China Postdoctoral Science Foundation. Correspondence and requests for materials should be addressed to X.-M.J or C.-Z.P.

**Author Contributions**

X.-M.J and C.-Z.P designed and supervised the whole project. X.-M.J, Z.-H.Y, B.Y, F.Z performed the experiment. T.Y and C.-Z.P designed the electric devices. X.-M.J analyzed the data and wrote the paper.

**Competing financial interests**

The authors declare that they have no competing financial interests.

**Figure Captions:**

**Figure 1. Schematic of experimental setup and principle diagram of EFT. a,** The configuration of EFT. SP is guided into UMZI$_1$ and its $V$ mode is delayed about 2.5 ns. The noisy channel is composed of a polarization controller (PC), 10m or 0.8km SMF, and a noise simulator (NS). UMZI$_2$ introduces a time delay of 2.5ns on the $H$ mode of SP. The probe laser (633nm) is coupled into and out of the EFT system by two dichromatic mirrors (DM, T808R633). **b,** principle diagram of EFT, the real line denotes the $V$ mode of SP and the dashed denotes the $H$ mode. **c,** Observed three peaks corresponding to the interval of $\Delta t$, $\Delta t + T$ and $\Delta t + 2T$. **d,** Observed envelope by measuring the 2-fold coincidence between TP and SP.

**Figure 2. The performance of EFT against BFE. a,** The result of EFT against BFE for three complementary bases, linear polarization state $|H\rangle, |+\rangle$ and circular polarization state

$|R\rangle$. **a, b, c,** The performance in the case of 10m SMF. **d, e, f,** The performance in the case of 0.8km SMF. The gray parties are the error rates simulated by NS. The red histograms show the error rates with EFT. The green panes show the efficiencies of transmission. Error bars are given by Poissonian statistics.

**Figure 3. The performance of the system against PSE and transmission of entanglement. a.** The stability of system against PSE. Experimental data is collected continuously in 1 hour with 0.8km fiber. **b.** The experimental measurements on S value in CHSH inequality. The Green histograms denote the measured S values without EFT system and the red histograms are the measured S values with EFT system. The dashed line marks the boundary of entanglement or not. S value above the line implies that entanglement is preserved with the assistant of EFT. Error bars are given by Poissonian statistics.

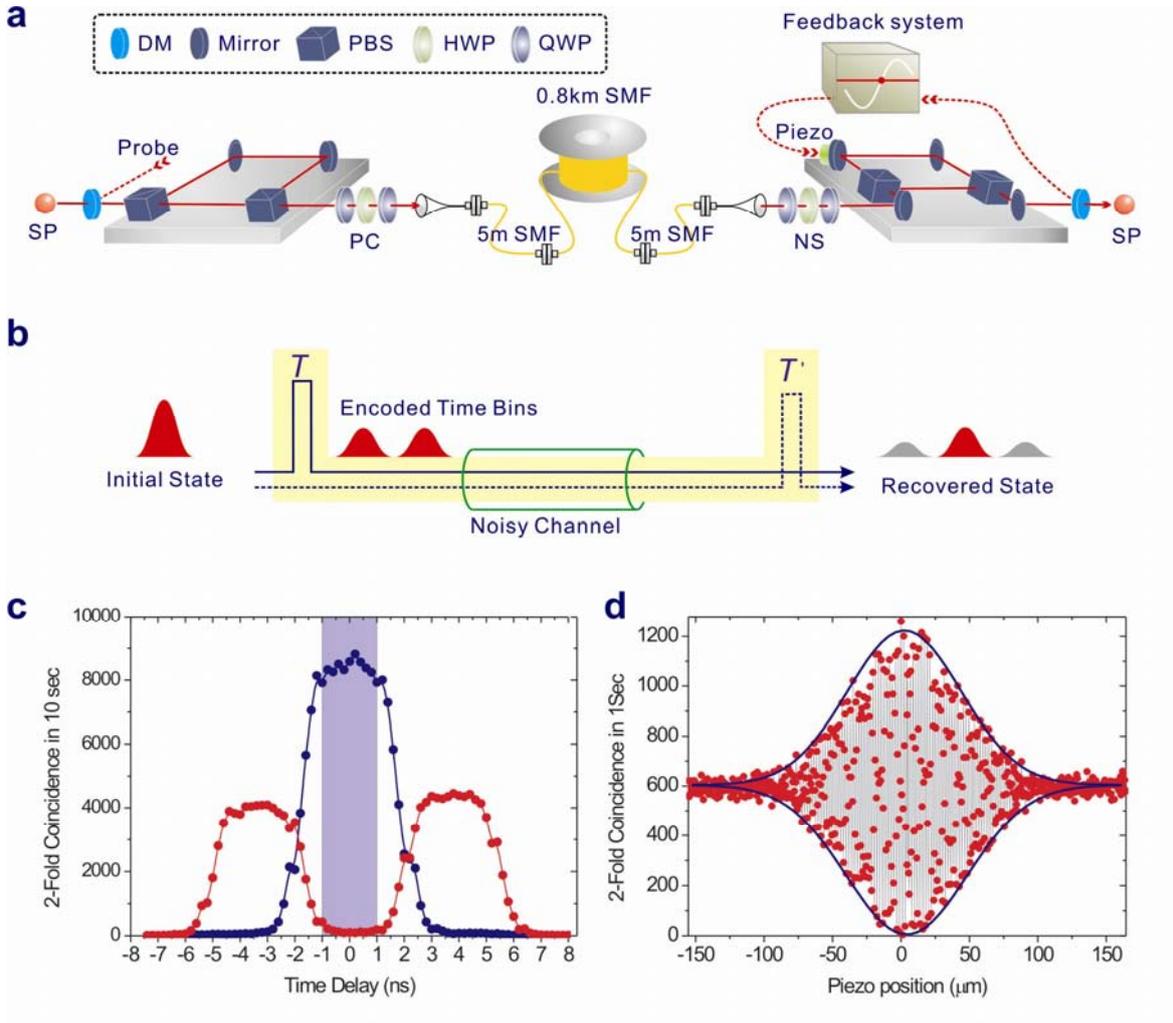

Figure-1

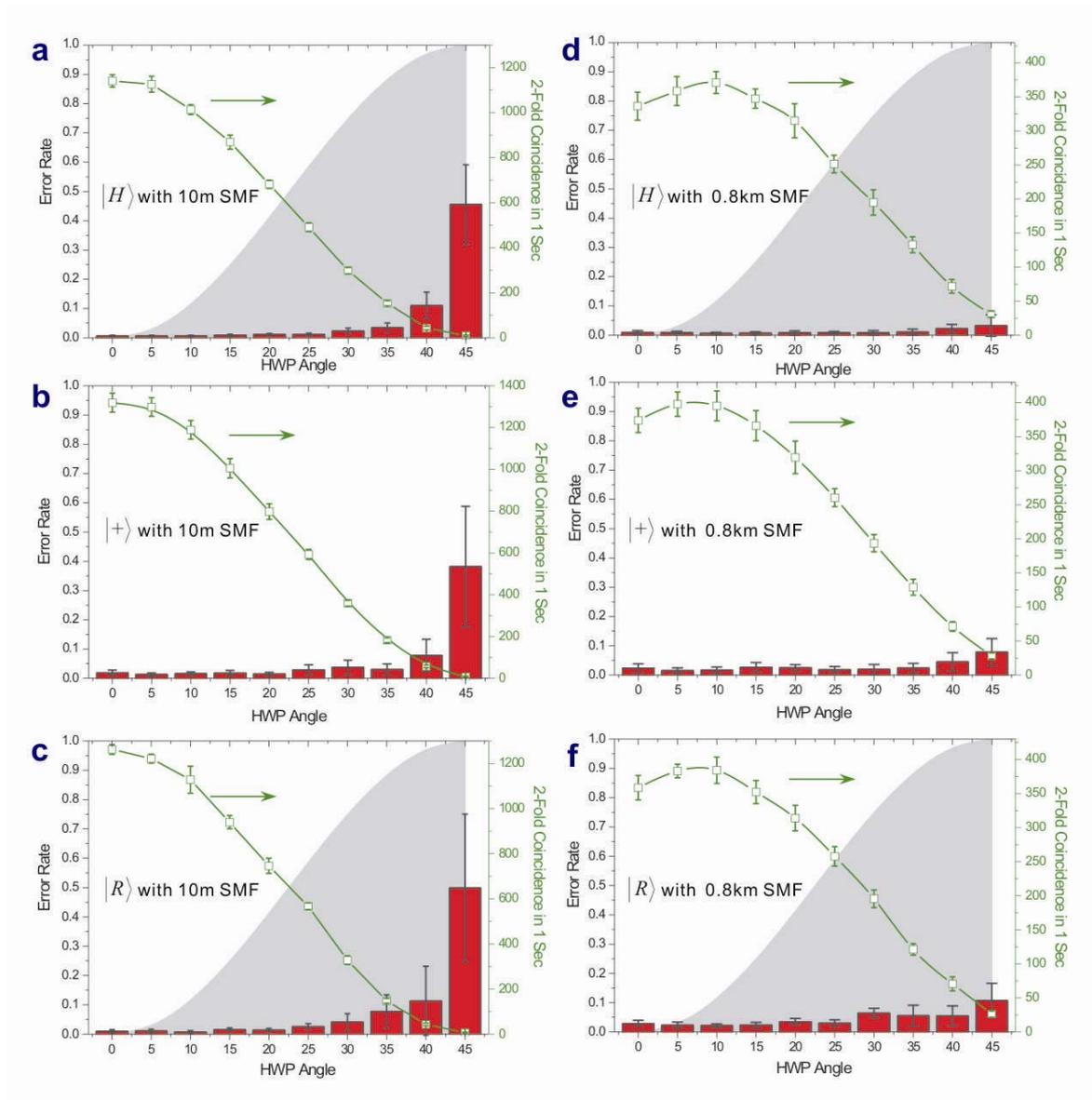

Figure-2

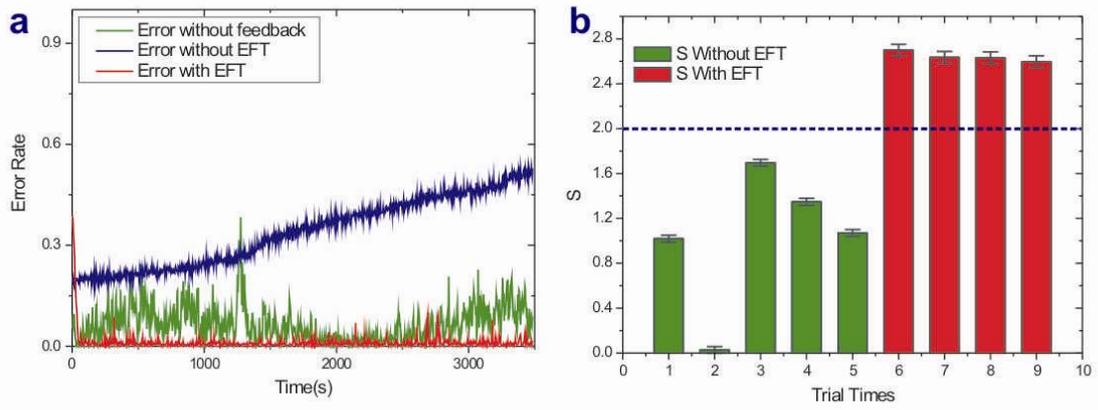

Figure-3